\documentclass[aps,nofootinbib,prd,twocolumn]{revtex4}
\usepackage[dvips]{graphicx}

\usepackage{amsmath}
\usepackage{amssymb}
\usepackage{caption}
\usepackage{hyperref}
\usepackage{upquote}
\usepackage{color}

\def\vev#1{\left\langle #1\right\rangle}
\def\SM{$\mathrm{SU(3)_c \otimes SU(2)_L \otimes U(1)_Y}$ }
\def\21{$\mathrm{SU(2)_L \otimes U(1)_Y}$ }

\newcommand{\AddrAHEP}{AHEP Group, Institut de F\'{i}sica Corpuscular --
  C.S.I.C./Universitat de Val\`{e}ncia, Parc Cientific de Paterna.\\
  C/Catedratico Jos\'e Beltr\'an, 2 E-46980 Paterna (Val\`{e}ncia) - SPAIN}

\newcommand{\Virginia}{Center for Neutrino Physics, Virginia Tech,
  Blacksburg, VA 24061, USA}

\newcommand{\Cinvestav}{Departamento de F\'{\i}sica, Centro de
  Investigaci{\'o}n y de Estudios Avanzados del IPN\\ Apdo. Postal
  14-740 07000 Mexico, DF, Mexico}
  
\begin{document}


\title{Constraining right-handed neutrinos}

\author{\underline{F. J. Escrihuela}~$^1$}\email{franesfe@alumni.uv.es}
\author{D. V. Forero~$^2$}\email{dvanegas@vt.edu}
\author{O. G. Miranda~$^3$}\email{omr@fis.cinvestav.mx}
\author{M. T\'ortola~$^1$}\email{mariam@ific.uv.es}
\author{J. W. F. Valle~$^1$} \email{valle@ific.uv.es, URL:
  http://astroparticles.es/} 
\affiliation{$^1$~\AddrAHEP}
\affiliation{$^2$~\Virginia}
\affiliation{$^3$~\Cinvestav}

\begin{abstract} 
  Several models of neutrino masses predict the existence of neutral
  heavy leptons. Here, we review current constraints on heavy
  neutrinos and apply a new formalism separating new physics from
  Standard Model. We discuss also the indirect effect of extra heavy
  neutrinos in oscillation experiments.
\end{abstract}

\maketitle
\section{Introduction}
\label{sec:introd}
In the Standard Model (SM), neutrinos are massless particles
contradicting the experimental observation of neutrino oscillations,
hence physics beyond the SM is required.

\begin{figure}
   \includegraphics[width=.8\linewidth]{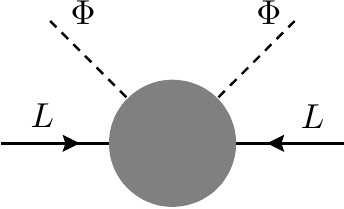}
    \caption{\label{fig:d5} 
    Dimension five  operator responsible for neutrino mass.}
\end{figure}

An effective lepton number violation dimension-five operator
$\mathcal{O}_5 \propto LL\Phi\Phi,$ can be added to the \SM model
in order to introduce neutrino masses
~\cite{weinberg:prd22, valle:jpc53}, where $L$ is one of the
three lepton doublets and $\Phi$ is the SM scalar doublet.

After electroweak symmetry breaking, Majorana neutrino masses are
induced, being proportional to $\vev{\Phi}^2$ and implying lepton
number violation. Hence, the smallness of neutrino mass, compared to
the masses of the SM charged fermions, arises from the smallness of
the coefficient in front of the operator $\mathcal{O}_5$ associated
with the lepton number violation by two units $(\Delta L = 2)$.

Unfortunately we cannot say too much more about this operator. We do
not have any clue about its mechanism, nor its mass scale, nor its
flavour structure.

A common possibility is to assume that $\mathcal{O}_5$ is induced, at
the tree level, by the exchange of heavy “messenger” particles. In
this way, seesaw models postulate neutral heavy states act as
“messenger” particles to induce neutrino mass. For instance, a
“right-handed” neutrino could be included associated to each of the
three isodoublet neutrinos (Type I seesaw)
~\cite{Valle:2015pba}:
\begin{equation}
m_{\nu} = \lambda_0 \frac{\vev{\Phi}^2}{M}
\end{equation}

The existence of processes with $\Delta L = 2$, such as neutrinoless
double beta decay, or lepton flavour violation processes (LFV) as $\,\mu
\to \,e \gamma$ , would give hints on the possible existence of these heavy Majorana neutrino messengers. Hence we could find signatures of heavy neutrinos and their mixings by studying this kind of processes.


\section{The method}

As we said in the previous section, heavy neutrinos are introduced in
several extensions of the SM such as linear and inverse seesaw models~\cite{mohapatra:prd34, concha:plb216, malinsky:prl95}, leading to a rich structure in the lepton mixing matrix. In
order to work with this kind of models we will use a symmetric
parameterization, consistent with the general formalism~\cite{schecter:prd22}, neatly separating ``new physics (NP)'' and ``Standard Model physics (SM)''.

For the case of three light neutrinos and $n –- 3$ extra heavy states, we
can construct the mixing matrix U as the product of $\omega_{ij}$
rotation matrices (Okubo's notation~\cite{schecter:prd22, okubo:ptp28}):
\begin{eqnarray}
U^{n \times m} &=& \omega_{n-1\, n}\:\omega_{n-2\,
  n}\:\ldots\:\omega_{1\, n} \nonumber \\
&&\omega_{n-2\, n-1}\:\omega_{n-3\, n-1}\:\ldots\:\omega_{2\,
  3}\:\omega_{1\, 3}\:\omega_{1\, 2}\label{eq:Untn}
\end{eqnarray}
\begin{equation}
\omega_{ij} = \begin{pmatrix} c_{ij} & 0 & e^{-i\phi_{ij}}s_{ij}\\
0 & 1 & 0\\
-e^{i\phi_{ij}}s_{ij} & 0 & c_{ij}
\end{pmatrix}\label{eq:om_matrix}
\end{equation}
where $c_{ij} = \cos\theta_{ij}$ and $s_{ij} = \sin\theta_{ij}$.

The mixing matrix $U$ can be decomposed in a new physics part and its
Standard Model part
\begin{eqnarray}
U^{n\times n} &=& R^{NP}\, R^{SM}\label{eq:Un_dec} \\
R^{NP} &=& \omega_{n-1\, n}\:\omega_{n-2\, n}\:\ldots\:\omega_{1\,
  n}\:\ldots\:\omega_{3\, 4}\:\omega_{2\, 4}\:\omega_{1\,
  4}\label{eq:R_NP} \\
R^{SM} &=& \omega_{2\, 3}\:\omega_{1\, 3}\:\omega_{1\,
  2}\label{eq:R_SM}
\end{eqnarray}
and it can be divided in four blocks
\begin{equation}
U=\begin{pmatrix}N & S\\
T & V
\end{pmatrix}\label{eq:ULindner}
\end{equation}
where $N$ is the block corresponding to the standard three neutrinos,
including their mixings between them and the extra neutrinos.

At the same time, the matrix $N$ can be also decomposed as
\begin{equation}
N=N^{NP}\, U^{SM}=\begin{pmatrix}\alpha_{11} & 0 & 0\\
\alpha_{21} & \alpha_{22} & 0\\
\alpha_{31} & \alpha_{32} & \alpha_{33}
\end{pmatrix}\: U^{SM}\label{eq:Ndescomp}
\end{equation}
where $U^{SM}$ is the usual mixing matrix of the Standard Model and
the matrix $N^{NP}$ includes all the new physics information through
the $\alpha_{ij}$ parameters (a more complete discussion is given in~\cite{paper3})
\begin{eqnarray}
\alpha_{11} &=& \: c_{1\, n}\: c_{\,1n-1}c_{1\, n-2}\ldots c_{14}\nonumber \\
\alpha_{22} &=& \: c_{2\, n}\: c_{\,2n-1}c_{2\, n-2}\ldots c_{24}
\nonumber \\
\alpha_{33} &=& \: c_{3\, n}\: c_{\,3n-1}c_{3\, n-2}\ldots c_{34}\label{eq:alphas} \\
\alpha_{21} &=& \:
    c_{2\, n}\: c_{\,2n-1}\ldots c_{2\, 5}\:\tilde{s}_{24}\bar{s}_{14}\nonumber \\  
    &+& c_{2\, n}\: \ldots c_{2\,
      6}\:\tilde{s}_{25}\bar{s}_{15}\:c_{14}\:+
    \tilde{s}_{2n}\bar{s}_{1n}\:c_{1n-1}\:c_{1n-2}\:\ldots\:c_{14} \nonumber  
\end{eqnarray}

In summary, by choosing a convenient order for the products of the
rotation matrices, $\omega_{ij}$, we can obtain a parameterization which puts all the information in a convenient form.

\subsection{Simplest extension of SM: 3 +1 neutrinos}

The formalism for the simplest extension of the SM includes one extra right handed singlet
\begin{equation}
\Psi_L = \begin{pmatrix}\nu_L\\ l_L\\
\end{pmatrix}\:\:\: , \:\:\: \mathcal{N}_R \label{eq:extranu}
\end{equation}
with the mixing relations between the gauge and mass eigenstates given as~\cite{mohapatra:book} 
\begin{equation}\label{eq:numix}
\nu_{k\, L} = \sum_{1}^{3} W_{k\, \alpha} \,\nu_{\alpha\, L} + S_{k\,
  4}\, \hat{\mathcal{N}}_{4\, L}
\end{equation}

The unitary mixing matrix $U^{4 \times 4}$ can be written as
\begin{equation}
U^{4\times4}=\begin{pmatrix}N^{3\times3} & S^{3\times1}\\
T^{1\times3} & V^{1\times1}
\end{pmatrix} = \begin{pmatrix} N_{e1} & N_{e2} & N_{e3} & S_{e4}\\
N_{\mu 1} & N_{\mu 2} & N_{\mu 3} & S_{\mu 4}\\
N_{\tau 1} & N_{\tau 2} & N_{\tau 3} & S_{\tau 4}\\
T_{41} & T_{42} & T_{43} & V\\
\end{pmatrix} \label{eq:U4} 
\end{equation}
where $N^{3\times3}$ is again the sub-matrix related with the standard
neutrinos Eq.(\ref{eq:Ndescomp})
It is important to notice that $N^{3 \times 3}$ is not unitary whereas
$U^{4 \times 4}$ is unitary because includes all neutrinos in the
model.

Comparing the terms in $N^{3 \times 3}$ with the terms of $U^{SM}$ we
obtain the following expressions for the $\alpha$ factors
\begin{eqnarray}
\alpha_{11} & = & c_{14}\nonumber \, ,\\
\alpha_{22} & = & c_{24}\nonumber \, , \\
\alpha_{33} & = & c_{34}\nonumber \, ,\\
\alpha_{21} & = & \tilde{s}_{24}\,\bar{s}_{14} \label{eq:alpha31} \, ,\\
\alpha_{32} & = & \tilde{s}_{34}\,\bar{s}_{24}\nonumber \, ,\\
\alpha_{31} & = & \tilde{s}_{34}\, c_{24}\, \bar{s}_{14}\nonumber\, .
\end{eqnarray}
\subsection{Aplication to 3 +3 model}

Usually, more than one extra neutrino is introduced in the theory, as
in sequential-type seesaw mechanisms where 3 (Type I) or 6 (Inverse
and Linear) extra singlets are included with the $SU(2)_L$ SM
doublets.

For such models our parameterization becomes 
\begin{equation}
U^{6\times 6}=\begin{pmatrix}N^{3\times 3} & S^{3\times 3}\\
T^{3\times 3} & V^{3\times 3}
\end{pmatrix}
\end{equation}
with these expressions for $\alpha$ the  parameters
\begin{eqnarray}
\alpha_{11}  &=&  c_{16}\, c_{15}\, c_{14}\nonumber \, ,\\
\alpha_{22}  &=&  c_{26}\, c_{25}\, c_{24}\nonumber \, ,\\
\alpha_{33}  &=&  c_{36}\, c_{35}\, c_{34} \nonumber \, ,\\
\alpha_{21}  &=&  \tilde{s}_{26}\,\bar{s}_{16}\, c_{15}\,c_{14} \; +
\; c_{26}\,\tilde{s}_{25}\,\bar{s}_{15}\, c_{14} \nonumber \; \\
&+& \; c_{26}\,c_{25}\,\tilde{s}_{24}\, \bar{s}_{14}  \label{eq:alfa_3+3} \, , \\
\alpha_{32}  &=&  c_{36}\,c_{35}\,\tilde{s}_{34}\,\bar{s}_{24} \; +
\; c_{36}\,c_{35}\,\bar{s}_{25}\, c_{24} \nonumber \; \\ 
&+& \; \tilde{s}_{36}\,\bar{s}_{26}\,c_{25}\, c_{24}  \nonumber \, ,\\
\alpha_{31} &=& c_{36}\, c_{35}\, c_{34} \, 
   \tilde{s}_{34}\,  c_{24} \, \bar{s}_{14} \;
+ \; c_{36}\, \tilde{s}_{35} \,  c_{24} \, \bar{s}_{15}\, c_{14}
\nonumber \; \\
&+& \; \tilde{s}_{36}\,  c_{26} \, \bar{s}_{16} \, c_{15}\, c_{14} 
 +  c_{36}\, \tilde{s}_{35}\, \bar{s}_{25} \, \tilde{s}_{24}\,
 \bar{s}_{14} \nonumber \; \\
&+& \; \tilde{s}_{36}\, \bar{s}_{26}\, c_{25} \, \tilde{s}_{24}\, \bar{s}_{14} \;
+ \; \tilde{s}_{36}\, \bar{s}_{26}\, \tilde{s}_{25} \, \bar{s}_{15}\,c_{14} \nonumber \, .
\end{eqnarray}
\section{Oscillation constraints}
The general expression for the survival and conversion neutrino
probability is given by~\cite{giunti:book}
\begin{eqnarray}
P_{\alpha \beta} &=& \delta_{\alpha \beta} - 4 \sum_{k>j}
\mathfrak{Re}\left[U^*_{\alpha k}\,U_{\beta k}\,U_{\alpha  j}\,U^*_{\beta j}\right]\, \sin^2\left(\frac{\Delta
    m^2_{kj}L}{4E}\right) \nonumber \\
     &+&   2 \sum_{k>j} \mathfrak{Im}\left(U^*_{\alpha k}\,U_{\beta k}\,U_{\alpha  j}\,U^*_{\beta j}\right] 
        \sin\left(\frac{\Delta m^2_{kj}L}{2E}\right) \label{eq:nuprobsm}
\end{eqnarray}
where $\delta_{\alpha \beta}$ appears due to the unitarity of the
mixing matrix. However, in a model with extra heavy neutrinos, this
equation will change because the mixing matrix
describing the three standard neutrinos ($N^{3 \times 3}$ in our
symmetric notation) will not be unitary and the effective probability
will not be normalized to 1. In this case, the probability includes
the $W$ terms from the truncated matrix $N^{3 \times 3}$
\begin{eqnarray}
P_{\alpha \beta} &=& \sum^3_{j,k} N^*_{\alpha k}\,N_{\beta
  k}\,N_{\alpha  j}\,N^*_{\beta j} \label{eq:nuprobnouni} \\
 &-& 4 \sum_{k>j}\mathfrak{Re}\left[N^*_{\alpha k}\,N_{\beta
  k}\,N_{\alpha  j}\,N^*_{\beta j}\right]\, \sin^2\left(\frac{\Delta
    m^2_{kj}L}{4E}\right) \nonumber \\
     &+&   2 \sum_{k>j} \mathfrak{Im}\left[N^*_{\alpha k}\,N_{\beta
  k}\,N_{\alpha  j}\,N^*_{\beta j}\right] 
        \sin\left(\frac{\Delta m^2_{kj}L}{2E}\right) \nonumber
\end{eqnarray}
For the electron (anti) neutrino survival probability, we get the expression  
\begin{eqnarray}
P_{ee} &=& \sum^3_{j} \left|N_{e j}\right|^2 \left|N_{e j}\right|^2 \nonumber \\
 &-& 4 \sum_{k>j} \left|N_{e k}\right|^2 \left|N_{e j}\right|^2 \, \sin^2\left(\frac{\Delta
    m^2_{kj}L}{4E}\right) \label{eq:eeprob}
\end{eqnarray}
and using Eq.(\ref{eq:numix}) and Eq.(\ref{eq:U4}) we obtain~\cite{paper3}
\begin{eqnarray}
P_{ee} &=& \alpha^4_{11} \bigg[ 1-\cos^4\theta_{13} \sin^22\theta_{12} \sin^2\left(\frac{\Delta
    m^2_{12}L}{4E}\right) \nonumber \\
&-& \sin^22\theta_{13} \sin^2\left(\frac{\Delta
    m^2_{13}L}{4E}\right) \bigg] \label{eq:eeprobalpha}
\end{eqnarray}
Note that the effect of the extra neutrinos is totally included in the
$\alpha^4_{11}$ factor, illustrating the utility of our symmetric
formalism.

Considering now only one extra fourth neutrino, Eq.(\ref{eq:eeprob}) would change to
\begin{equation}
P_{ee} \approx \cos^4\theta_{14} = \left(1-\left|S_{e4}\right|^2 \right)^2
\end{equation}

One could be tempted to study whether the presence of an extra light
singlet leptons (sterile neutrinos) could play some role in the
reported neutrino anomalies (as MiniBooNE~\cite{aguilar:prl110}). Unfortunately this is not the case and we
stick to the (natural) assumption that the extra states are heavy and
do not take part in oscillation effects. In this case one can get a constraint from some reported combined analysis~\cite{giunti:prd86}
\begin{eqnarray}
\sin^2\theta_{14} = \left|S_{e4}\right|^2 &<& 0.04 \:\: (90\%\, \text{C.L.}) \label{eq:sin14val} \\
\alpha^4_{11} = \cos^2\theta_{14} &<& 0.96 \label{eq:alpha11val}
\end{eqnarray}

\section{Future oscillation experiments}

As we can see from the previous section, it is not possible with
current experiments to obtain a very strong constraint on the new
physics $\alpha$ parameters so we consider future experimental
proposals such as LENA~\cite{wurn:lena}.

LENA is a future neutrino experiment which will use a $^{51}\text{Cr}$
artificial neutrino source with $5\, \text{MCi}$ intensity, producing a total
of $1.9 \times 10^5$ neutrino events. The expected number of neutrino
events for an energy recoil of the electron in the range from $200$ to
$550\, \text{keV}$ in the presence of an extra heavy neutrino would be given
by
\begin{equation}
N_{i} = \alpha^4_{11}\,n_e\,\phi_{Cr}\,\Delta t \int^{T_{i+t}}_{T_{i}}\int
\frac{d\sigma}{dT} R\left(T,\,T'\right)\,dT\,dT' \label{eq:lenaev}
\end{equation}
where $n_e$ is the number of electron targets, $\phi_{Cr}$ is the neutrino flux coming
from the source, $\Delta t$ is 28 days which corresponds to the
half-life of the source, and $R\left(T,\,T'\right)$ is the resolution
function
\begin{equation}
R\left(T,\,T'\right) = \frac{1}{\sigma \sqrt{2\pi}} \, \exp
\left[-\frac{(T-T')^2}{2\sigma^2}\right] \label{eq:lenaresfun}
\end{equation}
where $T$ is the recoil energy, $T'$ is the true energy and $\sigma =
0.075 \sqrt{TIMeV}$ is the expected energy resolution.

\begin{figure}
   \includegraphics[width=.9\linewidth]{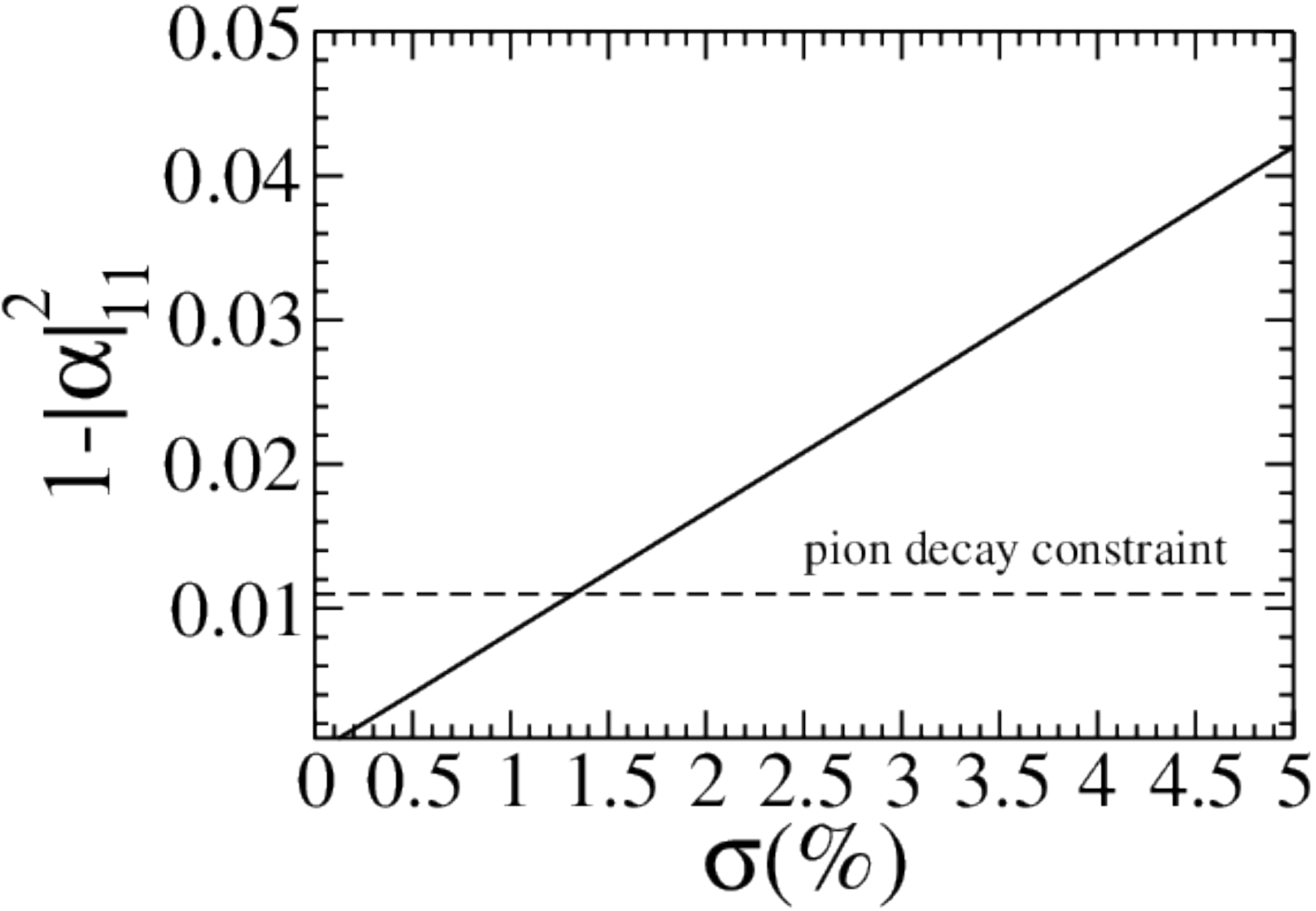}
    \caption{\label{fig:lena} 
   Estimation of LENA sensitivity.}
\end{figure}

An estimate of the expected sensitivity as a function of the total
percent error of the experiment can be performed in advance, being
shown in the Fig.~\ref{fig:lena}.

\section{Other constraints}

The effects of heavy neutrinos would show up as peaks in the leptonic
decays of pions and kaons or from their direct production at higher
collider energies~\cite{anupama:jhep05}. One can perform an analysis
of all these data, and combine the corresponding restrictions on heavy
neutrinos parameters.

\begin{figure*}
\begin{center}
   \includegraphics[width=.8\textwidth]{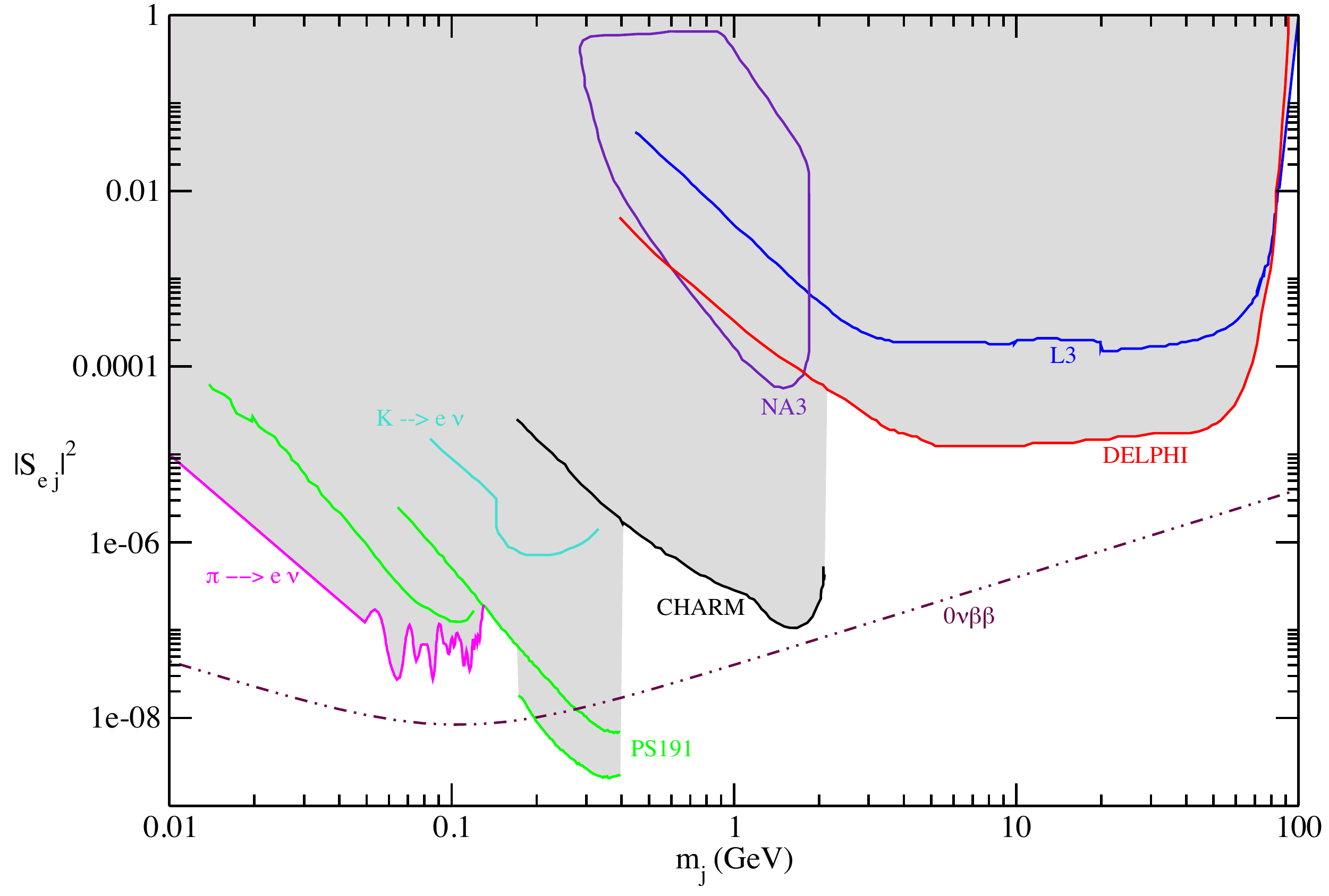}
    \caption{\label{fig:Sejplot} 
   Bounds on $\left|S_{ej}\right|^2$ versus the mass of extra heavy
   neutrino $m_{j}$. The $j$ in the labels correspond to the number of
   extra neutrinos $(j=4,5,6\ldots)$.}
\end{center}
\end{figure*}

\begin{figure*}
\begin{center}
\centering
   \includegraphics[width=1.0\textwidth]{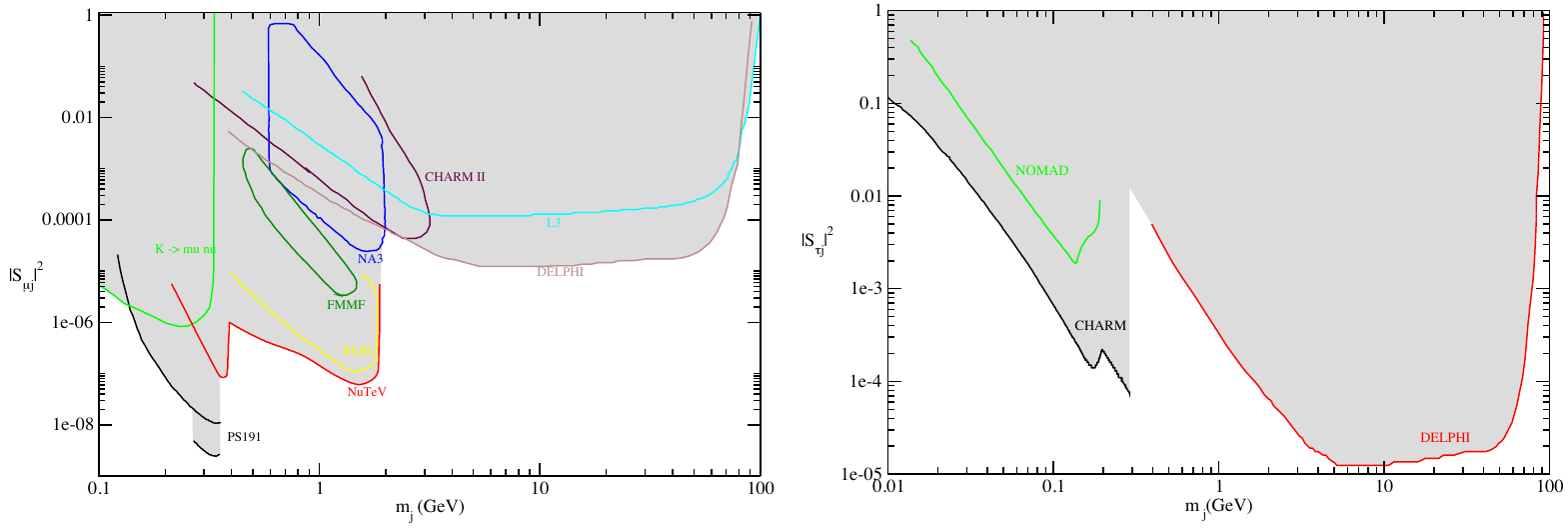}
    \caption{\label{fig:Smuj_Stauj} 
   \textit{Left panel}: Bounds on $\left|S_{\mu j}\right|^2$ versus the mass of extra heavy
   neutrino $m_{j}$. \textit{Right panel}: Bounds on $\left|S_{\tau j}\right|^2$ versus the mass of extra heavy
   neutrino $m_{j}$.}
\end{center}
\end{figure*}

Fig.~\ref{fig:Sejplot} compiles the bounds on the heavy neutrino
mixing with electron neutrinos at $90\,\%\, \text{C.L.}$
It agglutinates bounds coming from peak searches at lepton
decays, as $\pi \to e\,\nu$~\cite{britton:prl68, britton:prd46} and
$K \to e\, \nu$~\cite{berghofer:proceed, yamazaki:proceed}; meson
decays at PS191~\cite{bernardi:plb203}, NA3~\cite{badier:zpc31} and
CHARM~\cite{bergsma:plb166} and $Z^0$ decays at DELPHI~\cite{abreu:zpc74} and L3~\cite{adriani:plb295}. In
Fig.~\ref{fig:Sejplot} we show also the excluded region from
neutrinoless double beta decay experiments~\cite{benes:prd71}, valid
only if the heavy neutrinos are Majorana particles.
For completeness we show the bounds on the mixing with muon and tau
neutrinos in Fig.~\ref{fig:Smuj_Stauj}.
%



From a more complete combined analysis of these data we can get
constraints on our new physics parameters  associated to the presence
of the heavy neutrinos~\cite{paper3}.

\section{Conclusions}

Extra neutral heavy leptons are motivated in order to introduce
neutrino mass but no positive evidence of these particles has been
found so far.

Signatures of these heavy neutrinos arising from their mixings with
the light ones could be searched at laboratory experiments. The study
of these bounds would be useful to shed light upon the mass generation
mechanism of neutrinos and probe the scale of new physics. 

Models beyond SM, such as seesaw models, imply a very large number of
parameters. The symmetric parameterizatio of the neutrino mixing
matrix, describing the charged current, provides a very useful way to
separate new physics from SM effects, concentrating the information
and making easier to work with it. A more detailed account of our work
will be described elsewhere~\cite{paper3}.

\section{Acknowledgments}

FJE wants to thank the organizing committee for giving him the
opportunity to talk at ICHEP 14 and to his collaborators DVF, OGM, MAT
and JWFV for their advices. Likewise he thanks the AHEP group at
Valencia for hospitality.

This work has been supported by CONACyT grant 166639 (Mexico) and by the Spanish grants FPA2014-58183-P and Multidark CSD2009-00064 (MINECO), and PROMETEOII/2014/084 (Gen. Valenciana).


\begin{thebibliography}{1}

\bibitem{weinberg:prd22}S. Weinberg, Phys. Rev. {\bf D22}, 1694 (1980). 
\bibitem{valle:jpc53}J. W. F. Valle, J.Phys: Conf. Ser. {\bf 53}, 473
  (2006).
\bibitem{Valle:2015pba} Extensive discussion and references to the
  seesaw mechanism are given in J.~W.~F.~Valle and J.~C.~Romao, {\it
    Neutrinos in high energy and astroparticle physics}, (Wiley-VCH,
  Berlin 2015), 1st ed.
\bibitem{schecter:prd22}J. Schechter and J.W.F. Valle, Phys. Rev. {\bf
    D22}, 2227 (1980).
\bibitem{anupama:jhep05}A. Atre, T. Han, S. Pascoli and B. Zhang, JHEP {\bf 0905} 030 (2009). 
\bibitem{mohapatra:prd34}R. N. Mohapatra and J. W. F. Valle, Phys. Rev
  {\bf D34}, 1642 (1986).
\bibitem{concha:plb216}M. C. Gonzalez and J. W. F. Valle, Phys. Lett. {\bf B216}, 360 (1989).
\bibitem{malinsky:prl95}M. Malinsky, J. C. Romao and J. W. F. Valle,
  Phys. Rev. Lett. {\bf 95}, 161801 (2005).
\bibitem{okubo:ptp28}S. Okubo, Prog. Theor. Phys. {\bf 28}, 24 (1962) 
\bibitem{mohapatra:book}R. N. Mohapatra and P. B. Pal, Massive Neutrinos in Physics and Astrophysics (3rd ed.) (World Scientific, 2004). 
\bibitem{giunti:book}C. Giunti and C. H. Kim. Fundamentals of neutrino physics and astrophysics. (Oxford University Press, 2007).
\bibitem{aguilar:prl110}A. Aguilar-Arevalo et al. (MiniBooNE
  Collaboration), Phys. Rev. Lett. {\bf 110}, 161801 (2013).
\bibitem{giunti:prd86}C. Giunti, M. Laveder, Y. F. Li, Q. Y. Liu and
  H. W. Long, Phys. Rev. {\bf D86}, 113014 (2012).
\bibitem{wurn:lena}M. Wurn et al. (LENA Collaboration) (2011).
\bibitem{britton:prl68}D. I. Britton et al., Phys. Rev. Lett. {\bf 68}, 3000 (1992). 
\bibitem{britton:prd46}D. I. Britton et al., Phys. Rev. {\bf D46}, 885 (1992). 
\bibitem{berghofer:proceed}D. Berghofer et al., Proc. Intern. Conf. on
  Neutrino Physics and Astrophysics (Maui, Hawaii, 1981), 67 (1981),
  eds. R. J. Cence, E. Ma and A. Roberts, Vol. II (University of
  Hawaii, Honolulu, HI, 1981). 
\bibitem{yamazaki:proceed}T. Yamazaki, Proc. 22nd Intern. Conf, on High-energy physics (Leipzig, 1984), 262 (1984), eds. A. Meyer and E. Wieczorek, Vol. I (Akademie der Wiessenachaftender DDR, Leipzig, 1984).
\bibitem{bernardi:plb203}G. Bernardi et al., Phys. Lett. {\bf B203}, 332 (1988).
\bibitem{badier:zpc31}J. Badier et al. (NA3 Collaboration),
  Z. Phys. {\bf C31}, 21 (1986).
\bibitem{bergsma:plb166}F. Bergsma et al. (CHARM Collaboration),
  Phys. Lett. {\bf B166}, 473 (1986).
\bibitem{abreu:zpc74}P. Abreu et al. [DELPHI Collaboration],
  Z. Phys. {\bf C74}, 57 (1997) [Erratum-ibid. {\bf C75}, 580 (1997)].
\bibitem{adriani:plb295}O. Adriani et al. (L3 Collaboration),
  Phys. Lett. {\bf B295}, 371 (1992).
\bibitem{benes:prd71}P. Benes, A. Faessler, F. Simkovic and
  S. Kovalenko, Phys. Rev. {\bf D71}, 077901 (2005).
\bibitem{paper3}F.  J. Escrihuela, D. V. Forero, O. G. Miranda,
  M. T\'ortola and J. W. F. Valle, arXiv:1503.08879 [hep-ph].

\end{thebibliography}
\end{document}